\documentclass[prl,aps,twocolumn,10pt,showpacs,groupedaddress,superscriptaddress,floatfix,notitlepage]{revtex4-2}
\usepackage{amsmath,amsfonts,amssymb,graphics,graphicx,epsfig,color,times, braket}
\usepackage[utf8x]{inputenc}
\usepackage{color}
\usepackage{bbm, dsfont}
\usepackage{subfigure}
\usepackage{hyperref} 
\usepackage{mathrsfs}
\usepackage{verbatim} 
\begin{document}

\title{Adaptable Route to Fast Coherent State Transport via Bang-Bang-Bang Protocols}

\author{Ya-Tang~Yu}
\email{yatan1018@gmail.com}
\affiliation{Institute of Atomic and Molecular Sciences, Academia Sinica, Taipei 10617, Taiwan}

\author{Hsin-Lien~Lee}
\affiliation{Department of Physics and Center for Theoretical Physics, National Taiwan University, Taipei 10617, Taiwan}
\affiliation{Institute of Atomic and Molecular Sciences, Academia Sinica, Taipei 10617, Taiwan}

\author{Shao-Hung~Chung}
\affiliation{Institute of Atomic and Molecular Sciences, Academia Sinica, Taipei 10617, Taiwan}

\author{Ting~Hsu}
\affiliation{Department of Physics and Center for Theoretical Physics, National Taiwan University, Taipei 10617, Taiwan}

\author{Guin-Dar~Lin}
\affiliation{Department of Physics and Center for Theoretical Physics, National Taiwan University, Taipei 10617, Taiwan}
\affiliation{Center for Quantum Science and Engineering, National Taiwan University, Taipei 10617, Taiwan}
\affiliation{Trapped-Ion Quantum Computing Laboratory, Hon Hai Research Institute, Taipei 11492, Taiwan}
\affiliation{Physics Division, National Center for Theoretical Sciences, Taipei 10617, Taiwan}

\author{Ying-Cheng~Chen}
\affiliation{Institute of Atomic and Molecular Sciences, Academia Sinica, Taipei 10617, Taiwan}

\author{H.~H.~Jen}
\email{sappyjen@gmail.com}
\affiliation{Institute of Atomic and Molecular Sciences, Academia Sinica, Taipei 10617, Taiwan}
\affiliation{Physics Division, National Center for Theoretical Sciences, Taipei 10617, Taiwan}

\date{\today}
\renewcommand{\r}{\mathbf{r}}
\newcommand{\f}{\mathbf{f}}
\renewcommand{\k}{\mathbf{k}}
\def\p{\mathbf{p}}
\def\q{\mathbf{q}}
\def\bea{\begin{eqnarray}}
\def\eea{\end{eqnarray}}
\def\ba{\begin{array}}
\def\ea{\end{array}}
\def\bdm{\begin{displaymath}}
\def\edm{\end{displaymath}}
\pacs{}
\begin{abstract}
Fast coherent state transport is essential to quantum computation and quantum information processing. While an adiabatic transport of atomic qubits guarantees a high fidelity of the state preparation, it requires a long timescale that defies efficient quantum operations. Here, we propose an adaptable and fast bang-bang-bang (BBB) protocol, utilizing a combination of forward- and backward-moving trap potentials, to expedite the coherent state transport. We further showcase the advantage of applying squeezed coherent state evolution under a deeper potential followed by a weaker one, where a design of symmetric squeezing potential transports promotes an even shorter timescale for genuine state preparation. Our protocols outperform conventional forward-moving-only methods, providing new insights and opportunities for rapid state transport and preparation, ultimately advancing the capabilities of quantum control and quantum operations.
\end{abstract}
\maketitle
{\it Introduction}--Fast coherent control of quantum states is crucial for state manipulation and preparation, which is essential in applications of quantum computation and quantum information \cite{Nielsen2000, Cirac1995,Kielpinski2002, Leibfried2003, Haffner2008, Home2009, Lo2015, Bluvstein2022, Graham2022, Jandura2022, Moses2023, Evered2023, Chang2023, Bluvstein2024, Manetsch2025}. To maintain the coherence of quantum states \cite{Wineland1998,Parkins1999,Rowe2002,Reichle2006,Bowler2012,Odelin2019}, an adiabatic transport is sufficient to fulfill this goal, but it takes a time longer than needed for efficient operations in practical quantum platforms. Getting around of this obstacle, a recipe of shortcuts to adiabaticity \cite{Odelin2019, Chen2010, An2016, Kaufmann2018, Finzgar2025} provides fast routes to the coherent transport. This enables a rapid manipulation of quantum states—such as transporting a trapped particle—far faster than conventional adiabatic processes while still achieving the same high-fidelity final state. 

A substantial body of works has focused on similar non-adiabatic protocols for transferring a motional ground state with high fidelity at the destination \cite{Murphy2009, Lam2021, Hwang2025}, since this state minimizes thermal effects on internal-state operations \cite{Agarwal1971,Daniel1989,Wineland1998_1,Zhang2024}. This is even more important for trapped-ion systems to suppress errors in gate operations \cite{Cirac1995,Wineland1998_2}. The term `bang-bang' (BB) control has been widely investigated to accelerate this transport. While optimal-control approaches often uses `bang-bang' to describe piecewise-constant continuous dragging under relative dynamic constraints \cite{Xi2011, Torrontegui2011, Ding2020}, our work builds upon a distinct physical framework governed by absolute geometric bounds. Under these strict spatial constraints, BB control relies on instantaneous potential shifts executed much faster than the system's response time \cite{Alonso2013, Viola1999, Morton2006, Alonso2016, Damodarakurup2009}.

In this Letter, we propose an adaptable and fast bang-bang-bang (BBB) protocol, which facilitates the coherent state transport and shortens the timescale, surpassing the timescale limited by the BB protocol. In contrast to the forward-moving-only potential protocol, we demonstrate that a {\it backward} movement of the potential—counterintuitive to the transport direction—is necessary to further reduce the transport time in harmonic potentials. By designing a combination of forward- and backward-moving trap potentials in the BBB protocol, we showcase that this protocol can approach the idealized zero-time transport as long as an instantaneous potential shift is sufficed under a harmonic trap potential. Using realistic experimental parameters, our BBB protocol reduces the transport time by one third relative to the standard BB protocol while requiring only the initial and final potentials. For practical consideration when the spatial confinement of a trap is limited, we adopt a squeezed coherent state evolution \cite{Walls1983} projected on a deeper trap and manifest its advantage to further speed up the genuine state transport. While squeezed-BB (SBB) and single-frequency squeezed-BBB (SBBB) protocols naturally outpace the standard BB method by utilizing a higher trap frequency, SBBB yields the same transport time as SBB due to the intrinsic orientation constraints of the single-squeezed state. Notably, the key for extra improvement on the timescale lies at a weaker intermediate trap frequency in the double-squeezed BBB (DSBBB) protocol to relax this geometric constraint, enabling DSBBB to explicitly outperform SBBB. Our protocols outperform the conventional forward-moving-only potential methods, which offer unprecedented opportunities to fast state preparation and can be broadly implemented in both trapped-ion and neutral atom platforms.

{\it Model and phase-space dynamics}--We consider the transport of the motional state of a particle with mass $m$ in a one-dimensional harmonic trap. Initially ($t<0$), the particle is in the ground state $\ket{0_\text{init}}$ of the potential $V_\text{init}=m\omega_0^2\hat x^2/2$, where $\hat x$ is the position operator. The goal is to transport the particle a distance $d$ such that for $t>T$, it rests in the ground state $\ket{0_\text{dest}}$ of the destination potential, $V_\text{dest}=m\omega_0^2(\hat x-d)^2/2$ in a transporting time of $T$. We now look for a shortcut to the genuine state preparation by employing a moving harmonic potential with the same trap frequency $\omega_0$: $V_\text{move}(t)=m\omega_0^2[\hat  x-x_c(t)]^2/2$, with the trajectory of the potential's center denoted as $x_c(t)$. It must satisfy the boundary conditions $x_c(t< 0)=0$ and $x_c(t> T)=d$ to connect the initial and the destination potentials. The time-dependent Hamiltonian of the system can be expressed as 
\bea
H(t)=
\frac{\hat p^2}{2m}
+V_\text{move}(t),\label{H}
\eea
where $\hat p$ denotes a momentum operator. 

The ground state of a harmonic potential is a special coherent state with a minimal uncertainty in phase spaces ($\Delta x \Delta p=\hbar/2$ and $\Delta x (p)\equiv \hat x(\hat p) - \langle \hat x (\hat p)\rangle$), well-known for its shape-invariant wave-packet dynamics when it is displaced \cite{Sudarshan1963, Glauber1963, Zhang1990}. A coherent state $\ket{\alpha_0}$ can be generated by applying a displacement operator $\mathcal{D}(\alpha_0)$ to the ground state, $\ket{\alpha_0}=\mathcal{D}(\alpha_0)\ket{0}$, noting that $\alpha_0$ is complex in general and its imaginary part corresponds to a momentum displacement. In a stationary potential with trap frequency $\omega_0$, we define the dimensionless quadratures in the $\omega_0$ frame as $\hat X^{(0)}\equiv (\sqrt{m\omega_0/2\hbar})\hat x$ and $\hat P^{(0)}\equiv \hat p/\sqrt{2\hbar m\omega_0}$. For simplicity, we denote $\hat X^{(0)}$ and $\hat P^{(0)}$ as $\hat X$ and $\hat P$ in the following discussions unless otherwise specified. The complex amplitude can then be expressed as $\alpha = \langle \hat{X} \rangle + i \langle \hat{P} \rangle$, and evolves as $\alpha(t) = \alpha_0 e^{-i\omega_0 t}$. In this phase space representation \cite{Wigner1932, Scully1997}, the Wigner function of a coherent state is a circularly symmetric Gaussian, and its evolution corresponds to a clockwise rotation around the origin, which is the center of the potential \cite{SM}.

\begin{figure}[t]
	\centering
	\includegraphics[width=0.42\textwidth]{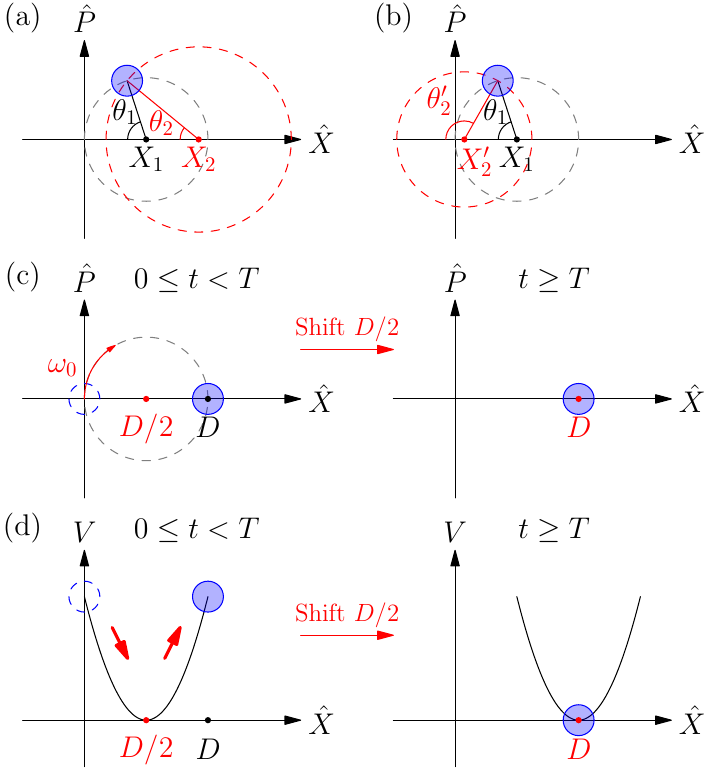}
	\caption{A comparison of potential shift directions, forward or backward, on the phase-space evolution. Starting from an initial state at an angle $\theta_1$ (when the potential is at $X_1$), (a) a forward potential shift to $X_2$ results in a new, smaller angle $\theta_2$. (b) By contrast, a backward shift to $X_2'$ results in a new, larger angle $\theta_2'$. This illustrates the key finding that $\theta_2 < \theta_1 < \theta_2'$, demonstrating that forward movements inhibit while backward movements accelerate the angular evolution toward $\pi$. (c) Phase-space evolution and (d) the corresponding potential trajectory for the standard BB protocol, which demonstrate the forward-moving speed limit $\tau_{\rm for}=\pi/\omega_0$. Dashed (filled) circles indicate the state of the particle at the beginning (end) of each free-evolution step.}\label{Fig1}
\end{figure}

{\it Role of the backward movement and the forward-moving minimal time bound}--For a successful transport, we constrain the state evolution by the boundary conditions as required in evolving Eq. (\ref{H}) toward the transport distance $D\equiv D^{(0)}\equiv(\sqrt{m\omega_0/2\hbar})d$ in a dimensionless form. First, the particle begins at rest ($\braket{X(0)}=\braket{P(0)}=0$), and the first potential move must be positive ($X_1 > 0$) to initiate the coherent state evolution along the positive direction. On the other hand, the final step must `catch' the state of the particle, bringing it to rest at the target ground state ($\braket{X(T)}=D, \braket{P(T)}=0$). This indicates that the state's center in phase spaces must evolve to the $X$-axis, without possessing any residual momentum. Therefore, the total accumulated angle of rotation during the transport is required to be a multiple of $\pi$, and the fastest process will correspond to the minimum angle, that is $\theta = \pi$. During this process, the momentum of the particle is always positive, and hence the final potential shift must also be a forward movement.

Notably, the opposite direction of the potential's movement affects this angular evolution in a favorable way. As shown in Fig. \ref{Fig1}(a), a positive potential shift reduces the accumulated angle ($\theta_2<\theta_1$), which prolongs the evolution toward $\pi$. In contrast, a backward potential movement, shown in Fig. \ref{Fig1}(b), increases the angle ($\theta_2'>\theta_1$) and accelerates the evolution due to $X_2'<X_1$. 

This insight immediately implies a forward-moving minimal time bound for any transport protocol that uses only forward-moving potential shifts. Since every positive step slows the angular evolution, the fastest protocol using purely forward-moving potentials must be the one that minimizes these steps. The optimal protocol is a two-step process, known as a BB protocol \cite{Alonso2016,Alonso2013} as illustrated in Figs. \ref{Fig1}(c,d). The potential is first shifted to $D/2$, displacing the particle and initiating the coherent state evolution. In a transport time $T$, the state fulfills the angle rotation of $\pi$ and reaches the $X$-axis. At this moment, a second and instantaneous shift from $D/2$ to $D$ effectively catches the particle at the destination without extra momentum ($\braket{P}=0$). Therefore, the time required for the transport of the BB protocol is $T = \pi/\omega_0$. This timescale represents the lower bound for any purely forward-moving transport: $\tau_{\rm for}(\omega_0) = \pi/\omega_0.$

\begin{figure}[t]
    \centering  \includegraphics[width=0.48\textwidth]{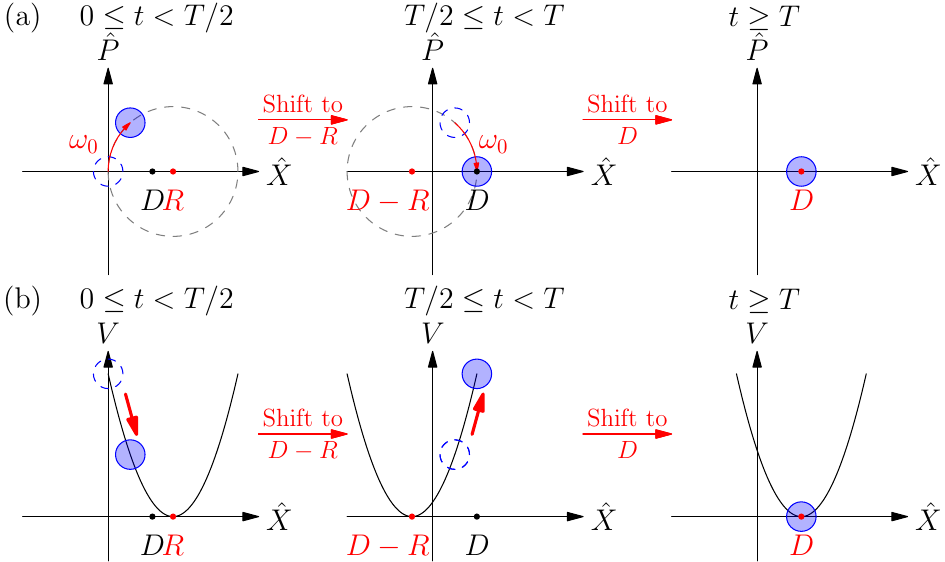}
    \caption{The schematic of the BBB protocol. (a) The evolution of a coherent state in the space under the displaced potential with its center at $R$ and $D-R$, respectively, in the first two bang processes of the BBB protocol, where the last bang process involves a shifted and holding potential at $D$. (b) The corresponding trajectory of the moving potential. The total transport time $T\equiv\tau_\text{BBB}(\omega_0, R)$. Dashed (filled) circles indicate the state at the beginning (end) of each free-evolution step.}\label{Fig2}
\end{figure}
{\it The BBB transport protocol}--To further reduce the timescale set by the BB protocol, we employ a backward movement in the second `bang' process of the BBB protocol in Fig. \ref{Fig2}, with the first and the last bang processes similar to the BB protocol. The trajectory of the center of the potential $X_{\rm c}^\text{BBB}$ is defined as 
\bea
\label{eq: BBBtrajectory}
X_c^\text{BBB}(R,t)=
\begin{cases}
0, & t<0\\
R, & 0\le t<\tau_\text{BBB}(R)/2\\
D-R, & \tau_\text{BBB}(R)/2\le t<\tau_\text{BBB}(R)\\
D, & t\ge \tau_\text{BBB}(R)
\end{cases},
\eea
where $\tau_\text{BBB}(R)$ indicates the total transport time, depending on the dimensionless displacement parameter $R\equiv R^{(0)}\equiv(\sqrt{m\omega_0/2\hbar})r$, where $r$ is the physical displacement of the potential's center. The protocol is designed to be symmetric with respect to the midpoint $D/2$. This symmetric construction is chosen to maximally utilize the available experimental control window while preserving the transport constraints \cite{SM}. At $t=0$, the potential is shifted to $R$. The state evolves until $t=\tau_\text{BBB}(R)/2$ when its position reaches $\braket{X}=D/2$. At this moment, a second bang is applied, shifting the potential to $D-R$, the position symmetric to $R$ with respect to the midpoint. The state continues to evolve, and due to this symmetry in displacements, it arrives at the destination $\braket{X}=D$ at $t=\tau_\text{BBB}(R)$ with zero momentum ($\braket{P}=0$). The final bang process at $t=\tau_\text{BBB}(R)$ shifts the potential to $D$, retaining the particle in the ground state of the destination potential.

The transport time for the BBB protocol can be calculated analytically, treating the trap frequency also as a parameter, which gives
\bea
\label{eq:BBBtime}
\tau_\text{BBB}(\omega_0,R)=\frac{2}{\omega_0}\cos^{-1}\left(\frac{2R-D}{2R}\right).
\eea
The above solution also contains the results of forward-moving scenarios when $R\le D/2$, where the case of $R=D/2$ reduces the protocol to the standard BB protocol, that is $\tau_\text{BBB}(\omega_0,D/2)=\pi/\omega_0$, recovering the timescale for $\tau_\text{for}$. Notably, for any $R > D/2$, the second step $D-R$ is less than $D/2$, constituting a backward movement. In this regime, the argument of the $\cos^{-1}$ in Eq. (\ref{eq:BBBtime}) becomes positive, and the transport time $\tau_\text{BBB}$ becomes shorter than $\tau_{\rm for}$. As $R$ increases, $\tau_\text{BBB}$ decreases monotonically, indicating that a larger controllable displacement enables faster transport. In experiments, however, $R$ cannot be chosen arbitrarily because the transport must remain within the effectively harmonic region of the trap \cite{SM}, where anharmonic corrections are negligible. In the idealized limit $R\rightarrow\infty$, one obtains $\lim_{R\rightarrow\infty}\tau_\text{BBB}(R)=0$; nevertheless, this does not violate physical principles because the approach to zero is still constrained by the fundamental quantum speed limits \cite{SM, Mandelstam1945, Margolus1998, Ness2022, Lam2021}.

\begin{figure*}[t]
    \centering
    \includegraphics[width=1\textwidth]{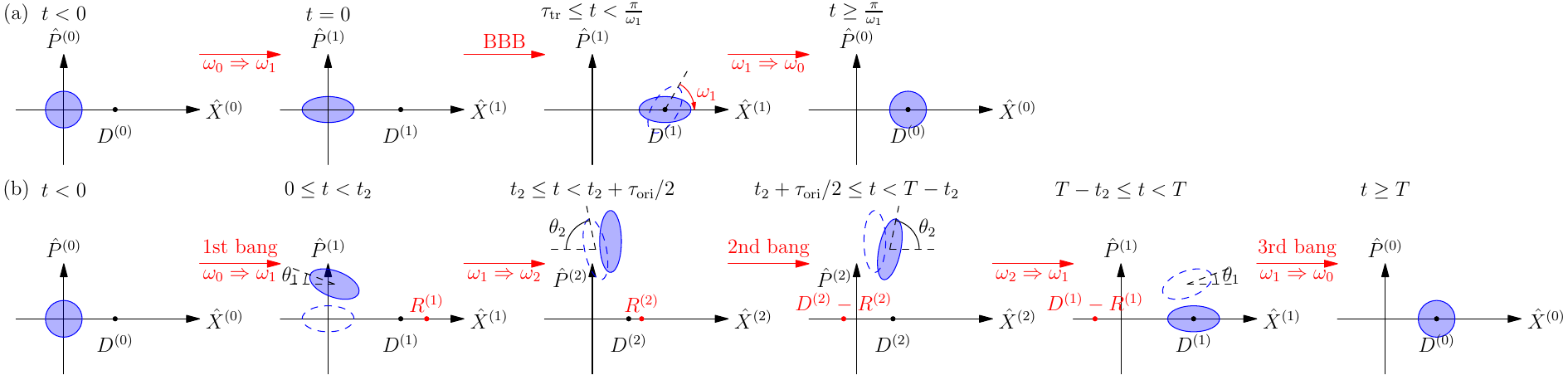}
    \caption{Schematic of two squeezed-BBB (SBBB) protocols. (a) A single-frequency SBBB protocol using $\omega_1$. While the spacial transport (time $\tau_{\rm tr}$) can be made faster, the final squeezed state's orientation is mismatched with the target, requiring an extra wait time $\tau_{\rm ex}$ before the final re-squeeze. (b) The double-squeezed-BBB (DSBBB) protocol. Under the second squeeze ($\omega_1 \to \omega_2$), the orientation angle becomes larger ($\theta_2 > \theta_1$) when projected onto the trap potential with a smaller $\omega_2<\omega_1$, accelerating the angular evolution. Dashed (filled) circles and ellipses indicate the state at the beginning (end) of each free-evolution step. (The dimensionless coordinates $\hat X^{(i)}$ and $\hat P^{(i)}$, and the parameters $R^{(i)}$ and $D^{(i)}$, evaluated in the $\omega_i$ frame, are defined in the text.)\label{Fig3}}
\end{figure*}
{\it The squeezed BBB (SBBB) protocols}--For practical considerations, there exists a maximum physical displacement parameter $r_{\rm max}$ that restricts the choice of parameter $R$. While the fundamental anharmonicity limits the effective confinement of the particle during the design phase, the post-fabrication reality of fixed hardware geometries—such as the static electrode structures in trapped-ion platforms—imposes an even stricter absolute constraint on the available spatial positions. Although the absolute displacement $R$ is strictly bounded by these instrumental limits, the trap depth—and thus its trap frequency $\omega$—remains highly controllable dynamically in systems like trapped ions and atoms in optical tweezers \cite{Beugnon2007}.

A simple and straightforward approach is to perform the entire BBB transport using a single, higher frequency $\omega_1 > \omega_0$. At $t=0$, one would switch $\omega_0 \to \omega_1$ and immediately begin the BBB protocol [see Fig. \ref{Fig3}(a)]. The frequency switching squeezes the state's Wigner function from a circle into an ellipse in the phase spaces. The spatial transport of the ellipse's center becomes faster, as $\tau_\text{BBB} \propto 1/\omega$. However, to prepare the target ground state at the end of the whole protocol, this scheme takes an extra time of $\tau_{\rm ex}$ for the ellipse rotating to the correct orientation as its major axis overlaps with the $X$-axis. The whole process including the transport time $\tau_{\rm tr}=\tau_\text{BBB}(\omega_1, R)$ and $\tau_{\rm ex}$ are summed to be $\pi/\omega_1$, the same as the forward-moving minimal time bound $\tau_{\rm for}(\omega_1)$. This indicates that the time for SBBB protocol is exactly the same as squeezed-BB (SBB) protocol when a single higher $\omega_1$ is applied.

To optimally utilize the spatial constraints and eliminate the extra time required for mismatched orientation, we propose the double-squeezed-BBB (DSBBB) protocol, as illustrated in Fig. \ref{Fig3}(b). To accurately track the state across different trap frequencies, we define the generalized dimensionless position, momentum, potential displacement, and destination in the $\omega_i$ frame as $\hat X^{(i)}\equiv (\sqrt{m\omega_i/2\hbar})\hat x$, $\hat P^{(i)}\equiv \hat p/\sqrt{2\hbar m\omega_i}$, $R^{(i)}\equiv (\sqrt{m\omega_i/2\hbar})r$, and $D^{(i)}\equiv (\sqrt{m\omega_i/2\hbar})d$. The key is to use a weaker intermediate trap frequency $\omega_2 < \omega_1$ to accelerate the overall ellipse rotations. The DSBBB protocol proceeds as follows:\\
(i) First bang and squeeze: At $t=t_1=0$, we apply the first bang by shifting the potential to $R^{(0)}$ and squeeze the state ($\omega_0 \to \omega_1$). The coordinate $R^{(0)}$ is concurrently transformed into $R^{(1)}$ under the $\omega_1$ frame. The state evolves under $\omega_1$ for a time $t_2$, accumulating an orientation angle $\theta_1 = \omega_1 t_2$.\\
(ii) Second squeeze: At $t=t_2$, we apply a second squeeze using a smaller trap frequency $\omega_2$ ($\omega_1 \to \omega_2$). This frequency change suddenly alters the orientation angle from $\theta_1$ to $\theta_2$ \cite{EM}. The system then evolves under $\omega_2$ for a time $\tau_\text{int}/2$, where $\tau_\text{int}=[\pi - 2\theta_2(t_2, \omega_1, \omega_2)]/\omega_2$ denotes the total duration of the intermediate stage under $\omega_2$. During this evolution, the ellipse's orientation angle reaches $\pi/2$, while the center of the state rotates about $R^{(2)}$ and arrives at half of the transport distance.\\
(iii) Second bang: At $t=t_2 + \tau_\text{int}/2$, the second bang is applied, shifting the potential to the symmetric position $D^{(2)} - R^{(2)}$. The state continues to evolve under $\omega_2$ for another duration of $\tau_\text{int}/2$, bringing the orientation angle to $\pi - \theta_2$.\\
(iv) Symmetric re-squeeze: At $t = t_2 + \tau_\text{int} \equiv T - t_2$, the process is reversed symmetrically. We squeeze the state back ($\omega_2 \to \omega_1$). The state then evolves for another time $t_2$.\\
(v) Final bang and final squeeze: At $t=T$, we restore the trap frequency ($\omega_1 \to \omega_0$) and apply the final bang by shifting the potential to the destination $D^{(0)}$. This leaves the particle perfectly at rest in the target ground state.

\begin{figure}[t]
    \centering
    \includegraphics[width=0.45\textwidth]{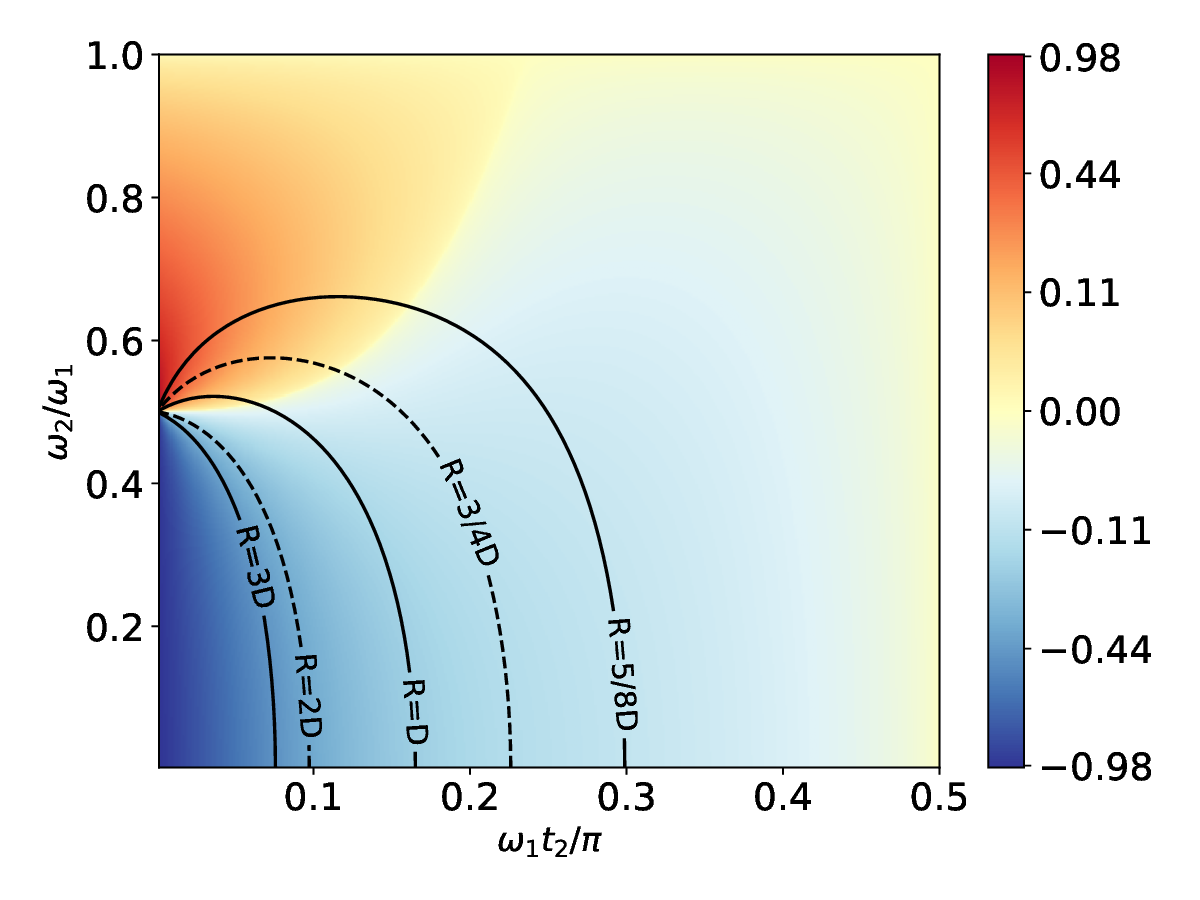}
    \caption{Phase diagram of the time advantage, $\tau_\text{DSBBB}-\pi/\omega_1$ (in units of $\pi/\omega_1$), for the parameters of intermediate trap frequency $\omega_2$ and the first-squeeze time $t_2$ at $\omega_1=2\omega_0$. The blue area shows where the DSBBB protocol evolves faster than a straightforward SBBB protocol with a single $\omega_1$. The solid and dashed contours indicate the required dimensionless displacement $R$ to realize the DSBBB protocol, where $R$ here denotes $R_\text{DSBBB}$ evaluated in the initial $\omega_0$ frame. The parameter space is plotted for $\omega_2$ in the region $(0,\omega_1]$ and $t_2$ in the region $(0,\pi/(2\omega_1)]$. The range of $t_2$ is bounded by $\pi/(2\omega_1)$ in a symmetrical process within the time $\pi/\omega_1$.}\label{Fig4}
\end{figure}

The total transport time for this DSBBB protocol is determined by the orientation evolution:
\bea
\tau_\text{DSBBB}(\omega_1,\omega_2, t_2)=2t_2+\frac{\pi-2\theta_2(t_2,\omega_1,\omega_2)}{\omega_2}.\label{DSBBB}
\eea
To ensure the state arrives exactly at the destination ground state, the required potential shift $R_{\rm DSBBB}$ \cite{EM} in the initial $\omega_0$ frame is given by
\bea
R_\text{DSBBB}(\omega_1,\omega_2, t_2) = \frac{D / 2}{1 - \cos\theta_1 \sin\theta_2 + \frac{\omega_1}{\omega_2} \sin\theta_1 \cos\theta_2}. \label{DSBBB_R}
\eea
In Fig. \ref{Fig4}, we plot the time advantage of $\tau_\text{DSBBB}$ over the single-frequency SBBB protocol ($\pi/\omega_1$). This shows that by choosing appropriate parameters $t_2$ and $\omega_2$ (blue area), $\tau_\text{DSBBB}$ can be made shorter than the simple SBBB protocol. The speedup is physically enabled by the slower intermediate frequency ($\omega_2 < \omega_1$), which induces a beneficial jump in the orientation angle projection ($\theta_2 > \theta_1$) and accelerates the overall rotation. To make the physical requirements explicit, we overlay contours of the required displacement parameter $R_\text{DSBBB}$ calculated from Eq. (\ref{DSBBB_R}). These contours indicate the minimum required spatial displacement needed to realize the DSBBB protocol, directly showing the experimentally accessible regions within the parameter space. Note that the idealized zero-time limit emerges again as $\omega_2$ and $t_2$ approach zero, due to the maximal jump of the orientation angle projection in the DSBBB protocol, which again relies on the assumption of no limit on displacement. For comparison, the standard BB protocol has a transport time of $\pi/\omega_0$. By utilizing a higher trap frequency ($\omega_1>\omega_0$), both the SBB and SBBB protocols reduce this time to $\pi/\omega_1$. For suitable choices of $\omega_2$ and $t_2$, the DSBBB protocol can reduce the time even further, thereby outperforming BB, SBB, and SBBB, because the second squeezing step relaxes the orientation constraint of the single-squeezed state.

{\it Experimental feasibility}--The physical constraint of the maximal potential displacement $r_{\rm max}$ is defined by the harmonic region of the potential, which is equivalent to the controllable strength of the coherent state under a fixed frequency. In trapped-ion platforms, it is already demonstrated that the BB control can be implemented to a coherent state with $|\alpha|\approx 100$, corresponding to displacing a $^{40}\text{Ca}^+$ ion by $1.49\mathrm{\mu}\text{m}$ with a trap frequency of $2.35\text{MHz}$ \cite{Alonso2016}. It is therefore promising that our proposed BBB protocol could be implemented in a similar experimental setup by controlling the harmonic potentials centered at $R$ and $D-R$ in Fig. \ref{Fig2}.
A particularly relevant case is $R=D$, where the BBB protocol uses only the initial and final potentials, unlike the standard BB protocol which also requires an intermediate one. Because these endpoint potentials must already be optimized for coherent state preparation, this choice minimizes calibration overhead. Assuming a similar allowable coherent state of $|\alpha| \approx 100$, our BBB protocol with $R=D$ could transport a $^{40}\text{Ca}^+$ ion over $0.785\ \mathrm{\mu}\text{m}$ in two thirds of the time required by the standard BB protocol.

When utilizing trap frequency as a control parameter for squeezed protocols, physical limitations establish strict upper and lower bounds. The upper bound ($\omega_{\max}$) is constrained by radio-frequency (RF) stability: the Mathieu quation coefficient $q$ must remain in the stable pseudo-potential regime (typically $q < 0.5$) \cite{Leibfried2003}. Pushing $\omega$ higher requires increasingly large RF voltages, leading to stronger micromotion, elevated RF heating, and breakdown risks. Conversely, the lower bound ($\omega_{\min}$) is set by electric-field noise, where anomalous heating rate $\dot n$ sharply increases at lower secular frequencies (typically $\dot{n} \propto 1/\omega^{\beta}$ with $1.9 \lesssim \beta \lesssim 2.4$), potentially overwhelming the coherent transport \cite{Brownnutt2015}. Furthermore, while the ideal protocol assumes instantaneous potential shifts, our analysis shows that this time advantage is highly robust against realistic drive limitations. Replacing the instantaneous jumps with continuous linear ramps of duration $\tau_{\rm sw}$ merely adds a single switching duration to the total transport time, preserving the fundamental speedup enabled by the backward motion \cite{SM}. In other platforms like optical tweezers \cite{Kaufman2012, Barredo2016, Barredo2018, Brown2019, Ebadi2021, Scholl2021, Dordevic2021}, the available controlled motional quanta might be lower (e.g., $|\alpha|^2 \approx 20$). The achievable distance for the BBB protocol, therefore, becomes smaller, but the general principle of accelerating transport by incorporating backward motion within a continuous trajectory still remains feasible.

{\it Conclusions and outlook}--In this work, we propose an advanced bang-bang-bang protocol that can outperform the conventional forward-moving potential protocols by utilizing backward potential movements. This provides an adaptable route to transport a particle to the desitination with high fidelity ground state. The achievable time advantage is set by the experimentally controllable harmonic range of the trap potential; using the parameters of Ref.~\cite{Alonso2016}, we showcase the BBB protocol reduces the transport time by one third compared to the standard BB protocol. Furthermore, we showcase the advantage of applying squeezed coherent state evolution under a deeper followed by a weaker potential. This double-squeezing approach relaxes intrinsic geometric constraints and further speeds up the state transport, surpassing the limitations set by the standard bang-bang-bang protocol. These protocols are promising for implementation in the trapped-ion and optical tweezer systems, providing a pathway to accelerating the speed of quantum computation and quantum information applications.

Extending our protocols to more realistic environments opens promising avenues for future research. Optimal quantum-control techniques \cite{Caneva2009, Furst2014} or machine-learning approaches \cite{Zhang2018} may be explored for transporting atoms at finite temperatures \cite{Pagano2024, Morandi2025}, where the trade-off between high-fidelity state preparation and transport time can be systematically investigated. While our baseline analysis shows that the protocol is robust against finite switching-time effects \cite{SM}, incorporating potential fluctuations from electrical or laser-field noise, along with more complex experimentally relevant pulse shapes \cite{Alonso2013, Cicali2025}, will help determine the ultimate limits of transport performance—crucial for scalable atom-array architectures used in quantum simulation and quantum computation. Finally, manipulating locally squeezed coherent states offers a means to generate continuous-variable entanglement \cite{Serafini2009} in multi-ion systems, providing an alternative route toward quantum information processing.

{\it Acknowledgments}--We acknowledge support from the National Science and Technology Council (NSTC), Taiwan, under the Grants No. 112-2112-M-001-079-MY3 and No. NSTC-114-2119-M-001-005, and from Academia Sinica under Grant AS-CDA-113-M04. We are also grateful for support from TG 1.2 of NCTS at NTU.  

{\it Data Availability}--The data are not publicly available. The data are available from the authors upon reasonable request. 

\begin{widetext}
\section*{End Matter}
\end{widetext}

\begin{figure*}[t]
	\centering
	\includegraphics[width=0.9\textwidth]{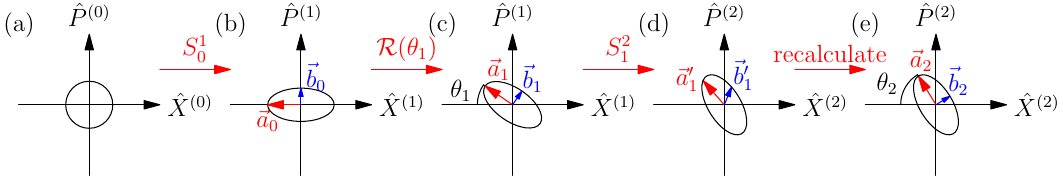}
	\caption{The transformations of the semi-major and semi-minor axes. (a) The particle is initially a standard coherent state, which is a circle in the $\hat X^{(0)}-\hat P^{(0)}$ phase space. (b) At $t=t_1=0$, a sudden change of trap frequency $\omega_0\to\omega_1$ squeezes the state into an ellipse in the $\hat X^{(1)}-\hat P^{(1)}$ phase space. (c) The ellipse rotates, accumulating an orientation angle of $\theta_1=\omega_1 t_2$. (d) At $t=t_2$, the frequency changes again ($\omega_1\to\omega_2$), but the transformed vectors $\vec{a}_1'$ and $\vec{b}_1'$ are not the new semi-major/minor axes. (e) The new principal axes are recalculated to compute $\theta_2$. The black circles/ellipses show the 1-sigma contour of the standard/squeezed coherent states.}\label{FigS3}
\end{figure*}
{\it The major axis orientation of the squeezed state}--When the trap frequency $\omega$ is changed, a coherent state is squeezed. The Wigner function of a state in a harmonic potential with frequency $\omega$ undergoes shape-preserving clockwise rotation in the appropriate phase space. Therefore, as introduced in the main text, correctly analyzing the dynamics requires a coordinate system scaled by the corresponding frequency. We recall the definitions of the dimensionless quadratures in the $\omega_i$ frame:
\bea
\hat X^{(i)}=(\sqrt{m\omega_i/2\hbar})\hat x 
\quad \text{and} 
\quad \hat P^{(i)}=\hat p/\sqrt{2\hbar m\omega_i}.
\eea

\noindent{\bf First Squeeze and Rotation}

Initially, a standard coherent state is prepared in the harmonic potential with trap frequency $\omega_0$. Its Wigner function is a circularly symmetric Gaussian in the $\hat X^{(0)}-\hat P^{(0)}$ phase space [Fig. \ref{FigS3}(a)].\\
At $t=t_1=0$, the frequency suddenly changes to $\omega_1$. The coordinates are transformed from the $\hat X^{(0)}-\hat P^{(0)}$ system to the $\hat X^{(1)}-\hat P^{(1)}$ system by the matrix
\bea
\label{eq: transformation_01}
S_0^1=
\begin{pmatrix}
\sqrt{\omega_1/\omega_0} &0\\
0 & \sqrt{\omega_0/\omega_1}
\end{pmatrix}.
\eea

This transforms the initial circle into an ellipse [Fig. \ref{FigS3}(b)]. If $\omega_1>\omega_0$, the ellipse is `flat' (squeezed in momentum); if $\omega_1<\omega_0$, it is `thin' (squeezed in position). We define the initial principal semi-axis vectors (which correspond to the 1-sigma uncertainties) as
\bea
\vec{a}_0 = (\tfrac{1}{2}\sqrt{\omega_1/\omega_0}, 0)^T\quad \text{and}\quad  \vec{b}_0 = (0, \tfrac{1}{2}\sqrt{\omega_0/\omega_1})^T.
\eea
Their lengths represent the uncertainties $|\vec{a}_0|\equiv\Delta X^{(1)}$ and $|\vec{b}_0|\equiv\Delta P^{(1)}$, which fulfill the minimum uncertainty relation $\Delta X^{(1)} \Delta P^{(1)}=1/4$. As time evolves, the ellipse rotates clockwise at frequency $\omega_1$. At time $t_2$, the accumulated angle is $\theta_1=\omega_1 (t_2-t_1)$ [Fig. \ref{FigS3}(c)]. The two axis vectors are transformed by the clockwise rotation matrix $\mathcal{R}(\theta_1)$ as
\bea
\vec{a}_1=\mathcal{R}(\theta_1)\vec{a}_0 \quad \text{and} \quad \vec{b}_1=\mathcal{R}(\theta_1)\vec{b}_0,
\eea
where
\bea
\label{eq: rotation_matrix}
\mathcal{R}(\theta_1)=
\begin{pmatrix}
\cos \theta_1 & \sin \theta_1\\
-\sin \theta_1 & \cos \theta_1
\end{pmatrix}.
\eea

\noindent{\bf Second Squeeze and New Axis Calculation}

At $t=t_2$, the second squeeze is performed ($\omega_1 \to \omega_2$). We map the state into the $\hat X^{(2)}-\hat P^{(2)}$ coordinate system using the matrix
\bea
\label{eq: transformation1-2}
S_1^2=
\begin{pmatrix}
\sqrt{\omega_2/\omega_1} & 0\\
0 & \sqrt{\omega_1/\omega_2}
\end{pmatrix}.
\eea

Applying this transform to the rotated axes gives $\vec{a}_1'=S_1^2\vec{a}_1$ and $\vec{b}_1'=S_1^2\vec{b}_1$ [Fig. \ref{FigS3}(d)]. A problem occurs: because $S_1^2$ is a squeezing operation (not a simple rotation), these new vectors $\vec{a}_1'$ and $\vec{b}_1'$ are no longer orthogonal and do not represent the semi-major/minor axes of the new ellipse. We must find the new principal axes. The points on the new ellipse (which was the 1-sigma contour) can be parametrized by $\phi$
\bea
\label{eq: ellipse}
\vec{v}(\phi)=\cos(\phi)\vec{a}_1'+\sin(\phi)\vec{b}_1'.
\eea

The semi-major and semi-minor axes correspond to the extrema of the distance $|\vec{v}(\phi)|$. These occur at angles $\alpha$ and $\beta$ satisfying $d/d\phi\,[|\vec{v}(\phi)|^2]_{\phi=\alpha,\beta}=0$, with $\beta=\alpha+\pi/2$. This calculation yields the condition for the extrema angles
\bea
\tan(2\alpha)=\frac{2\vec{a}_1'\cdot\vec{b}_1'}{|\vec{a}_1'|^2-|\vec{b}_1'|^2}.
\eea

Substituting the two resulting angles into Eq. (\ref{eq: ellipse}) gives the new semi-major and semi-minor axis vectors, $\vec{a}_2$ and $\vec{b}_2$
\bea
\vec{a}_2=\cos(\alpha)\vec{a'}_1+\sin(\alpha)\vec{b'}_1 \\
\vec{b}_2=\cos(\beta)\vec{a'}_1+\sin(\beta)\vec{b'}_1.
\eea
Finally, we can calculate the new orientation angle $\theta_2$ of the semi-major axis $\vec{a}_2$ in the $\hat X^2-\hat P^2$ coordinate system [Fig. \ref{FigS3}(e)]. If the new major axis vector is $\vec{a}_2 = (a_2^x, a_2^p)$, the angle is given by:
\bea
\theta_2=\tan^{-1}(-a_2^p/a_2^x).
\eea

{\it Derivation of the required potential shift for the DSBBB protocol}--To give an analytical form of the required potential shift $R_\text{DSBBB}$ for the DSBBB protocol and formalize the physical sequence illustrated in Fig. \ref{Fig3}(b) of the main text, we track the center of the state in the comoving frame, defining the coordinates as $X^{(i)}_\text{com} \equiv X^{(i)} - X_c^{(i)}$ and $P^{(i)}_\text{com} \equiv P^{(i)}$. After the first bang, the potential is shifted to $X_c^{(1)} = R^{(1)}_\text{DSBBB}$ in the $\omega_1$ frame. At $t=0$, the state is at the origin, so its position in the comoving frame is $(-R^{(1)}_\text{DSBBB}, 0)$. 

After evolving for a time $t_2$, it rotates clockwise by an angle $\theta_1 = \omega_1 t_2$. The coordinates in the comoving frame at $t=t_2$ can be obtained by applying the rotation matrix defined in Eq.~(\ref{eq: rotation_matrix}):
\bea
X_\text{com}^{(1)}(t_2) &=& - \cos\theta_1 \cdot R^{(1)}_\text{DSBBB}, \\
P_\text{com}^{(1)}(t_2) &=& \sin\theta_1 \cdot R^{(1)}_\text{DSBBB} .
\eea

At $t=t_2$, the trap frequency is changed instantaneously from $\omega_1$ to $\omega_2$. To transform these comoving coordinates into the $\omega_2$ frame, we apply the squeeze matrix defined by Eq.~(\ref{eq: transformation1-2}). Using the geometric relation $R^{(2)}_\text{DSBBB} = \sqrt{\omega_2/\omega_1} \cdot R^{(1)}_\text{DSBBB}$, the transformed comoving coordinates are:
\bea
X_\text{com}^{(2)}(t_2) &=& - \cos\theta_1 \cdot R^{(2)}_\text{DSBBB}, \\
P_\text{com}^{(2)}(t_2) &=& \frac{\omega_1}{\omega_2}\sin\theta_1 \cdot R^{(2)}_\text{DSBBB}.
\eea

Before the second bang, the state orbits the fixed trap center ($X_{c}^{(2)} = R^{(2)}_\text{DSBBB}$) for half of the intermediate time, $\tau_\text{int}/2$. The orbital angle swept during this period is $\omega_2 \tau_\text{int}/2 = \pi/2 - \theta_2$. Rotating the comoving coordinates clockwise by this angle yields the comoving position at the midpoint of the protocol, and the corresponding absolute position is:
\bea
X^{(2)}(t_2+\tau_\text{int}/2) = X^{(2)}_\text{com}(t_2+\tau_\text{int}/2)+R^{(2)}_\text{DSBBB} \nonumber\\
= \left( 1 - \cos\theta_1 \sin\theta_2 + \frac{\omega_1}{\omega_2} \sin\theta_1 \cos\theta_2 \right) \cdot R^{(2)}_\text{DSBBB}.
\eea

For a valid symmetric protocol, the state must arrive at the halfway point of the transport distance exactly at half of the total transport time. This requires the absolute position to be $D^{(2)}/2$:
\bea
(1 - \cos\theta_1 \sin\theta_2 + \frac{\omega_1}{\omega_2} \sin\theta_1 \cos\theta_2) \cdot R^{(2)}_\text{DSBBB}= \frac{D^{(2)}}{2}.
\eea

Finally, we project this requirement back into the initial $\omega_0$ frame. Recalling that $R^{(2)}_\text{DSBBB} = \sqrt{\omega_2/\omega_0} \cdot R^{(0)}_\text{DSBBB}$ and $D^{(2)} = \sqrt{\omega_2/\omega_0} \cdot D^{(0)}$, the frequency scaling factors perfectly cancel out. Dropping the $(0)$ superscript for simplicity (such that $R_\text{DSBBB} \equiv R^{(0)}_\text{DSBBB}$ and $D \equiv D^{(0)}$), we isolate $R_\text{DSBBB}$, which exactly yields Eq.~(\ref{DSBBB_R}) in the main text. This derives the exact analytical form of the initial potential shift required to perform the DSBBB protocol, which successfully catches the particle at the destination in the ground state.

\newpage
\begin{widetext}
\section{Supplementary Information:\\ Adaptable Route to Fast Coherent State Transport via Bang–Bang–Bang Protocols}


\begin{figure*}[t]
	\centering
	\includegraphics[width=0.9\textwidth]{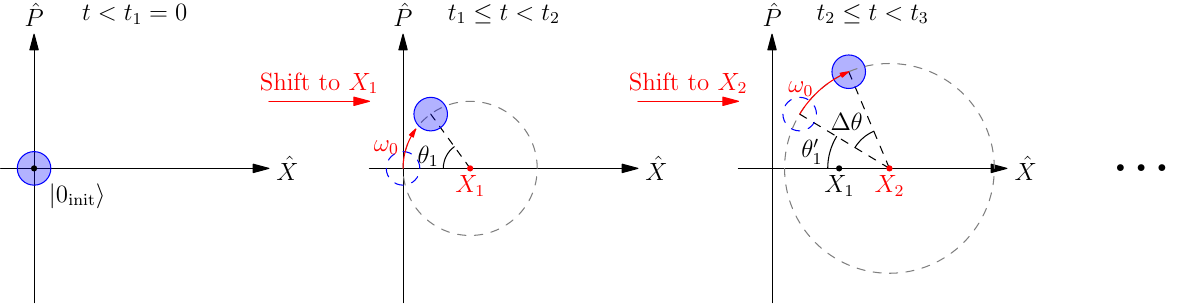}
	\caption{Phase-space evolution of a coherent state under an $N$-step moving potential. The state begins at the origin (the ground state $\ket{0_{\rm init}}$). At $t_1=0$, the potential shifts to $X_1$, causing the state to rotate around $(X_1, 0)$ until $t_2$. At $t_2$, the potential shifts to $X_2$, and the state's center of rotation switches to $(X_2, 0)$, continuing until $t_3$. This process repeats for all subsequent steps until $t=T$. Dashed (filled) circles indicate the coherent state at the beginning (end) of each free-evolution step.}\label{FigS1}
\end{figure*}
\section{Coherent State Evolution under a Stepwise Potential Trajectory}
We consider an $N$-step trajectory $X_c^N(t)$ for a moving harmonic potential. The potential traps a particle of mass $m$ with a constant trap frequency $\omega_0$. The trajectory, expressed in the dimensionless $\hat X-\hat P$ coordinate system, is given by:
\bea
\label{eq: trajectory}
X_c^N(t)=
\begin{cases}
0, & t<t_1=0\\
X_1, & t_1\le t<t_2\\
\vdots & \\
X_i, & t_{i}\le t<t_{i+1}\\
\vdots & \\
X_{N-1}, & t_{N-1}\le t<t_N\\
D, & t\ge t_N=T
\end{cases},
\eea
The dimensionless coordinates are defined as $\hat X\equiv(\sqrt{m\omega_0/2\hbar})\hat x$ (position) and $\hat P\equiv\hat p/ \sqrt{2\hbar m\omega_0}$ (momentum), where $\hat x$ and $\hat p$ are the position and momentum operators. The total transport distance $d$ is represented by the dimensionless quantity $D=(\sqrt{m\omega_0/2\hbar})d$. In the trajectory, $X_i$ is the potential's center at step $i$, $t_i$ is the time of the $i$-th potential shift, and $T=t_N$ is the total transport time.

The evolution of a coherent state in phase space under this kind of discrete potential trajectory consists of a series of clockwise rotations. We visualize the first two steps in Fig. \ref{FigS1}, and the process is described as follows:\\
(i) Initially, the particle begins at the ground state of the initial potential, corresponding to the initial condition $(\braket{X(0)} = \braket{P(0)} = 0)$.\\
(ii) At $t=t_1=0$, the potential suddenly shifts to $X_1$. The coherent state then begins to rotate clockwise around the point $(X_1,0)$ with angular frequency $\omega_0$.\\
(iii) The rotation continues until $t=t_2$, accumulating a phase angle $\theta_1=\omega_0 t_2$.\\
(iv) At $t=t_2$, the potential shifts again to $X_2$. The state's center of rotation instantly switches to $(X_2, 0)$, and it begins to rotate around this new point.\\
(v) The rotation continues until $t=t_3$, accumulating a further angle $\Delta\theta=\omega_0(t_3-t_2)$.\\
(vi)  This process is repeated for all subsequent shifts. At the final step ($t=t_N=T$), the potential shifts to $D$ and remains there, `catching' the particle at the destination as the ground state.

\begin{figure*}[t]
	\centering
	\includegraphics[width=0.6\textwidth]{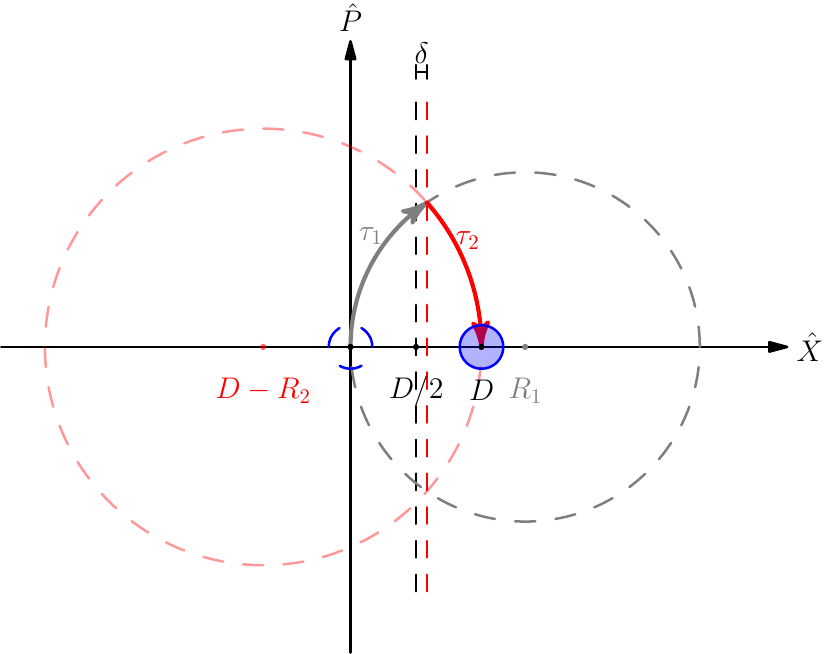}
	\caption{Schematic of the asymmetric BBB protocol. A valid transport can be constructed by solving for the common chord of two circles centered at $R_1$ and $D-R_2$.}\label{FigS_Asym_Schematic}
\end{figure*}

\subsection{Asymmetric Bang-Bang-Bang Protocol Analysis}
\label{app:AsymmetricBBB}

In the main text, we focus on the symmetric BBB protocol because it represents the time-optimal solution when utilizing a fixed amount of experimental resources, specifically the spatial extent of the harmonic trap. To demonstrate this optimality, we analyze a generalized asymmetric BBB protocol here.

Suppose the trap maintains a good harmonic approximation up to a maximum displacement of $R_\text{max}$. We can construct an asymmetric protocol by applying a first potential shift to $R_1$ for a duration $\tau_1$, and a second shift to position $D - R_2$ for a duration $\tau_2$. The trajectory of the potential's center is defined as:
\bea
X_c(t)=
\begin{cases}
0, & t<0\\
R_1, & 0\le t<\tau_1\\
D-R_2, & \tau_1\le t<\tau\\
D, & t\ge \tau,
\end{cases}
\eea
where $\tau=\tau_1+\tau_2$ is the total transport time.

By solving for the boundary conditions required for a valid state transport that connects the initial and destination ground states (illustrated in Fig. \ref{FigS_Asym_Schematic}), we find the required durations of the two waiting periods to be:
\bea
\omega_0\tau_1=\cos^{-1}\left(\frac{2R_1-D-2\delta}{2R_1}\right)
\eea 
and 
\bea
\omega_0\tau_2=\cos^{-1}\left(\frac{2R_2-D+2\delta}{2R_2}\right)
\eea
where $\delta=\frac{(R_2-R_1)D}{2(R_1+R_2-D)}$. 

Consequently, the total transport time as a function of the two displacements is given by:
\bea
\omega_0\tau(R_1,R_2)=\cos^{-1}\left(\frac{2R_1-D-2\delta}{2R_1}\right)+\cos^{-1}\left(\frac{2R_2-D+2\delta}{2R_2}\right).
\eea

This analytical solution allows us to directly compare symmetric and asymmetric strategies. For symmetric cases where $R_1 = R_2 = R$, the parameter $\delta$ vanishes, and the total time reduces to the standard $\tau_\text{BBB}$ given in the main text. In this regime, the transport time monotonically increases as $R$ decreases, meaning $\tau(R,R) > \tau(R_\text{max}, R_\text{max})$ for all $R < R_\text{max}$.

\begin{figure*}[t]
	\centering
	\includegraphics[width=0.6\textwidth]{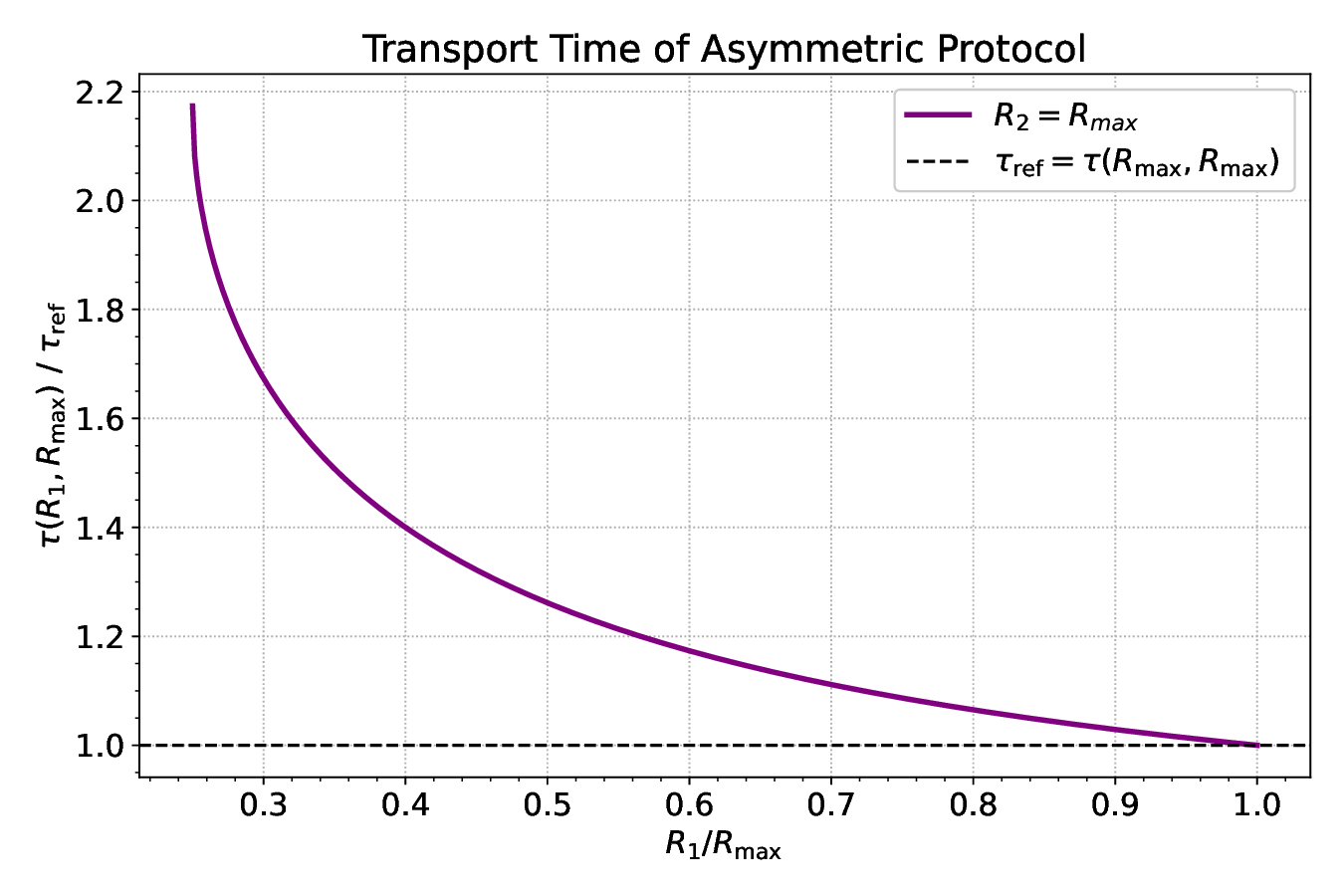}
	\caption{Transport time optimization for the asymmetric protocol. Total transport time $\tau(R_1, R_2)$ is plotted versus the intermediate displacement $R_1$, with $R_2$ fixed at the maximum allowable displacement $R_\text{max}$ ($R_\text{max}=2D$). The transport time is strictly minimized when the protocol is symmetric ($R_1 = R_\text{max}$).}\label{FigS_Asym_Time}
\end{figure*}

To address asymmetric configurations where $R_1 \neq R_2$, we examine the case where one parameter is maximized to the experimental limit ($R_2 = R_\text{max}$) while the other varies. As demonstrated in Fig. \ref{FigS_Asym_Time}, the transport time is strictly minimized when $R_1$ approaches $R_\text{max}$. Due to the symmetry of the equation itself [$\tau(R_1, R_2) = \tau(R_2, R_1)$], this result holds regardless of which step is restricted. 

Therefore, assuming the spatial constraints of the trap are identical throughout the entire transport process, a symmetric protocol maximally utilizes the available linear harmonic range and yields the shortest possible transport time. If experimental limitations dictate unequal spatial constraints on either side of the transport, an asymmetric protocol can be constructed using these generalized equations to achieve optimal transport within those specific bounds.

\subsection{Anharmonic Corrections at Large Displacements}
\label{app:Anharmonicity}

In the ideal bang-bang-bang (BBB) protocol, the transport time decreases monotonically as the displacement parameter $R$ increases. However, at large displacement amplitudes, anharmonic corrections to the trapping potential become non-negligible. These corrections induce an amplitude-dependent frequency shift that alters the phase-space rotation rate from a constant $\omega_0$, directly modifying the geometric timing arguments underlying the protocol.

To quantify this effect, we model the trap potential as a harmonic term with a weak quartic correction, $V(x) = \frac{1}{2}m\omega_0^2 x^2 + \lambda x^4$, and treat the quartic term perturbatively. To first order in the perturbation strength $\lambda$, the frequency correction is linear in $\lambda$. In contrast, the correction to the final-state fidelity (which is calculated from the state probability) is of order $\mathcal{O}(\lambda^2)$. Therefore, we treat the fidelity as effectively invariant and focus on the primary dynamic consequence: the correction to the transport time.

In the ideal harmonic case, the dimensionless equations of motion for the center of the particle's wave packet in a moving potential are $\dot{X} = \omega_0 P$ and $\dot{P} = \omega_0 [X_c(t) - X]$. To incorporate the anharmonicity, we absorb the quartic perturbation into a position-dependent effective frequency, defined relative to the instantaneous trap center $X_c(t)$:
\bea
\omega\big(X, X_c(t)\big) = \omega_0 \sqrt{1 + 2\Lambda [X - X_c(t)]^2},
\eea
where $\Lambda = 2\lambda \hbar / (m^2 \omega_0^3)$ is the dimensionless anharmonicity strength. The modified general equations of motion then become:
\bea
\dot{X} &=& \omega\big(X, X_c(t)\big) P, \\
\dot{P} &=& \omega\big(X, X_c(t)\big) [X_c(t) - X].
\eea

For a symmetric process, the duration between the first and second bangs is identical to that between the second and third bangs, which we denote as $T_{\rm half}$. Before the second bang, the potential is held stationary at $X_c(0\le t<T_{\rm half}) = R$. Taking the ratio of these coupled equations under this stationary condition yields $dP/dX = (R - X)/P$. Integrating this relation gives $P^2 + (X - R)^2 = \text{const}$, which demonstrates that the phase-space trajectory remains a perfect circle. The anharmonicity therefore preserves the geometric path but alters the angular velocity along that path.

Parameterizing the trajectory with an angle $\theta$ such that $X - X_c(t) = R \sin\theta$, the angular velocity is:
\bea
\dot{\theta} = \omega_0 \sqrt{1 + 2\Lambda R^2 \sin^2\theta}.
\eea
By inverting this relation, we can express $T_\text{half}$ as an exact integral over the angular sweep from an initial angle $0$ to the required final angle $\theta_f = \cos^{-1}\left(\frac{2R-D}{2R}\right)$:
\bea
T_\text{half} = \frac{1}{\omega_0} \int_{0}^{\theta_f} \frac{d\theta}{\sqrt{1 + 2\Lambda R^2 \sin^2\theta}}.
\eea

Assuming the anharmonic correction is small such that $\Lambda R^2 \ll 1$, we can Taylor expand the integrand as $(1 + 2\Lambda R^2 \sin^2\theta)^{-1/2} \approx 1 - \Lambda R^2 \sin^2\theta$. Evaluating the integral gives the corrected total transport time $T \equiv 2 T_\text{half}$:
\bea
T \approx \frac{2\theta_f}{\omega_0} - \frac{\Lambda R^2}{\omega_0} \left( \theta_f - \frac{\sin(2\theta_f)}{2} \right).
\eea

The first term is exactly the ideal BBB transport time $\tau_\text{BBB}(\omega_0, R)$. The second term reveals that the transport-time correction scales linearly with the dimensionless anharmonicity strength $\Lambda$. To recover the particle in the target ground state without residual excitation, the waiting time for each potential shift must be adjusted by this correction term to compensate for the anharmonicity-induced delay ($\Lambda<0$) or speed-up ($\Lambda>0$).


\subsection{Quantum Speed Limit Analysis of the Bang-Bang-Bang Protocol}
The minimal time for a quantum evolution is constrained by the Quantum Speed Limit (QSL). Two of the most fundamental bounds are those proposed by Mandelstam and Tamm (MT) \cite{Mandelstam1945} and Margolus and Levitin (ML) \cite{Margolus1998}. For a transition between an initial state $\ket{\psi_\text{init}}$ and a target state $\ket{\psi_\text{targ}}$, these bounds are given by
\bea
\tau_{\rm MT}
= \frac{\hbar}{\Delta E} \cos^{-1}\left(|\braket{\psi_{\rm init}|\psi_{\rm targ}}|\right),\\
\tau_{\rm ML}
= \frac{\hbar}{E} \cos^{-1}\left(|\braket{\psi_{\rm init}|\psi_{\rm targ}}|\right),
\eea
where $E$ is the time averaged mean energy (relative to the ground state), and $\Delta E$ is the time averaged energy spread (standard deviation).

For a transport process, $\ket{\psi_\text{init}}$ is the ground state of the initial potential, and $\ket{\psi_\text{targ}}$ is the ground state $\ket{\psi_\text{dest}}$ of the destination potential, which is displaced by a distance $d$. The overlap between these two harmonic oscillator ground states (for a particle of mass $m$ at frequency $\omega_0$) is
\bea
|\braket{\psi_{\rm init}|\psi_{\rm dest}}|
=\exp\left[-\frac{m\omega_0 d^2}{4\hbar}\right]=\exp\left[-\frac{D^2}{2}\right].
\eea
For a realistic transport, we assume the dimensionless distance $D$ is large enough (e.g., $D \approx 3$, where the overlap is around $1\%$) that the states are nearly orthogonal, which gives $\cos^{-1}(0) \approx \pi/2$.

For our Bang-Bang-Bang (BBB) protocol, the coherent state excitation is set by the parameter $R$, which corresponds to the initial dimensionless potential displacement used in the protocol. This results in a coherent state of magnitude $|\alpha|=R$, and for the whole BBB process, this magnitude remains invariant. For a coherent state in a quantum harmonic oscillator, the mean energy (relative to the ground state energy) is given by
\bea
E\equiv\braket{H}-E_0=\hbar\omega_0|\alpha|^2,\eea
and the energy spread (the standard deviation of energy) is
\bea\Delta E\equiv(\braket{H^2}-\braket{H}^2)^{1/2}=\hbar\omega_0|\alpha|,
\eea
where $H$ is the Hamiltonian of the harmonic oscillator. Substituting $|\alpha|=R$ into the QSLs with the $\pi/2$ orthogonality approximation, we find the bounds for the BBB protocol, which are
\bea
\tau_{\rm MT}^{\rm BBB}(R)\approx\frac{\pi}{2R\omega_0},\\
\tau_{\rm ML}^{\rm BBB}(R)\approx\frac{\pi}{2R^2\omega_0}.
\eea

The transport time of the BBB protocol (at frequency $\omega_0$) is
\bea
\label{eq:BBBtime}
\tau_{\rm BBB}(R)=\frac{2}{\omega_0}\cos^{-1}\left(1-\frac{D}{2R}\right).
\eea

In Fig. \ref{FigS2}, we plot these three timescales as a function of $R$. $\tau_{\rm BBB}$ clearly remains above both QSL bounds for finite $R$. In the limit $R\rightarrow\infty$, we can expand the inverse cosine function in Eq. \ref{eq:BBBtime} into a Puiseux series, where
\bea
\cos^{-1}(1-x)=\sqrt{2x}\left( 1 + \frac{1}{12}x + \mathcal{O}(x^2) \right).
\eea
This gives the asymptotic behavior of the BBB time as
\bea
\tau_{\rm BBB}(R \rightarrow \infty) \approx \frac{2}{\omega_0}\sqrt{2\left(\frac{D}{2R}\right)} = \frac{2\sqrt{D}}{\omega_0 \sqrt{R}}.\eea
Since $\sqrt{R}$ scales slower than $R$, our protocol time $\tau_{\rm BBB} \propto 1/\sqrt{R}$ is well above the MT bound $\tau_{\rm MT} \propto 1/R$ and the ML bound $\tau_{\rm ML} \propto 1/(R)^2$ for large $R$.

Our analysis confirms that $\tau_{\rm BBB}$ does not violate these two fundamental QSLs. This result suggests that either (i) an even faster protocol may exist that more closely approaches these limits, or (ii) a different QSL, one more specific to transport protocols, could provide a tighter bound for this setup.
\begin{figure*}[t]
	\centering
	\includegraphics[width=0.5\textwidth]{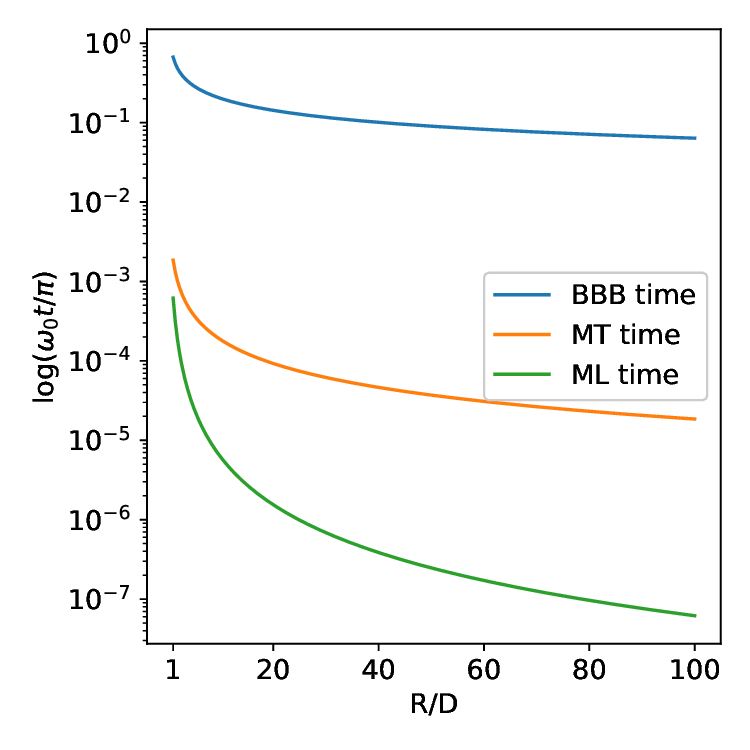}
	\caption{Timescales with respect to the parameter $R$ over the range $[D,100D]$. The blue line is the BBB transport time $\tau_{\rm BBB}$, while the orange and green lines show the MT bound $\tau_{\rm MT}$ and the ML bound $\tau_{\rm ML}$, respectively. For finite $R$, $\tau_{\rm BBB}$ is always larger than $\tau_{\rm MT}$ and $\tau_{\rm ML}$, which shows that BBB protocol does not violate the fundamental principles of quantum mechanics.}\label{FigS2}
\end{figure*}
\newpage

\begin{figure*}[t]
\centering
\includegraphics[width=1\textwidth]{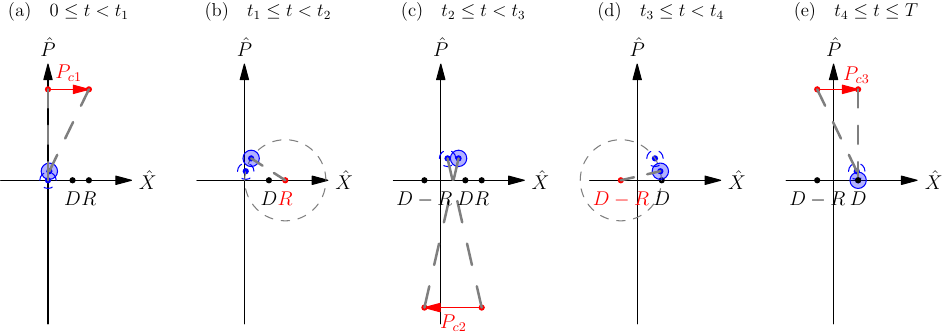}
\caption{Phase-space evolution for the BBB protocol with finite switching time. The instantaneous potential shifts are replaced by linear ramps of duration $\tau_{\rm sw}$, interspersed with stationary waiting intervals of $\tau_{\rm wait}$. The panels illustrate the state evolution sequentially: (a) The state undergoes circular motion in the comoving frame during the initial linear ramp to position $R$ with momentum $P_{c1}=R/\theta_{\rm sw}$. (b) The state rotates around the stationary trap center $(R,0)$ during the first wait period. (c) The potential linearly ramps backward to $D-R$  with momentum $P_{c2}=(D-2R)/\theta_{\rm sw}$, fulfilling the symmetry condition required for transport. (d) The state rotates around the stationary trap center $(D-R,0)$ during the second wait period. (e) The final linear ramp with momentum $P_{c3}=R/\theta_{\rm sw}$ brings the potential and the state to rest at the destination $D$.}\label{FigS_FiniteSwitching}
\end{figure*}

\subsection{Finite Switching Time Analysis}
\label{app:FiniteSwitching}

To evaluate the robustness of the time advantage of Bang-Bang-Bang (BBB) protocol under realistic experimental limitations, we model the instantaneous potential shifts as continuous linear ramps of finite duration $\tau_{\rm sw}$, during which the potential moves at a constant velocity. Between these shifts, the potential remains stationary for a hold time of $\tau_{\rm wait}$.

To describe the alternating moving (ramp) and stationary (wait) phases, we define the time markers for the trajectory as $t_1 = \tau_{\rm sw}$, $t_2 = \tau_{\rm sw} + \tau_{\rm wait}$, $t_3 = 2\tau_{\rm sw} + \tau_{\rm wait}$, $t_4 = 2\tau_{\rm sw} + 2\tau_{\rm wait}$, and the total transport time $T = 3\tau_{\rm sw} + 2\tau_{\rm wait}$. The full piecewise trajectory of the potential's center, $X_c(t)$, is then explicitly defined as:
\bea
X_c(t)=
\begin{cases}
\frac{R}{\tau_{\rm sw}}t, & 0\le t< t_1 \\
R, & t_1\le t<t_2 \\
R+\frac{D-2R}{\tau_{\rm sw}}(t-t_2), & t_2\le t<t_3 \\
D-R, & t_3\le t<t_4 \\
D-R+\frac{R}{\tau_{\rm sw}}(t-t_4), & t_4\le t\le T.
\end{cases}
\eea

This alternating sequence of translations and rotations in phase space is illustrated step by step in Fig. \ref{FigS_FiniteSwitching}(a)--(e). To determine the required hold time $\tau_{\rm wait}$ for valid ground-state-to-ground-state transport, we analyze the corresponding phase-space dynamics. For conciseness, we define the associated phase angles as $\theta_{\rm sw} = \omega_0 \tau_{\rm sw}$ and $\theta_{\rm wait} = \omega_0 \tau_{\rm wait}$.\\

\noindent{\bf Phase-Space Dynamics during a Linear Ramp}

Before analyzing the step-by-step evolution, we first establish that the state continuously undergoes pure circular motion in the comoving frame during a linear potential ramp. In the phase space, the equations of motion for the state's center are $\dot{X} = \omega_0 P$ and $\dot{P} = \omega_0 [X_c(t) - X]$.

Assuming a linear ramp potential, the trajectory of the trap center is $X_c(t) = X_{\rm start} + Vt$, where $V$ is a constant velocity. Taking the time derivative of the first equation of motion and substituting the second equation into it yields a second-order non-homogeneous differential equation for the position:
\bea
\ddot{X} + \omega_0^2 X = \omega_0^2 (X_{\rm start} + Vt).
\eea

The general solution to this equation consists of a homogeneous oscillatory part and a particular linear part:
\bea
X(t) = X_{\rm start} + Vt + A \cos(\omega_0 t) + B \sin(\omega_0 t),
\eea
where $A$ and $B$ are constants determined by the initial conditions. The corresponding momentum is obtained by differentiating the position, $P(t) = \dot{X}/\omega_0$:
\bea
P(t) = \frac{V}{\omega_0} - A \sin(\omega_0 t) + B \cos(\omega_0 t).
\eea

To clearly observe the circular motion, we define the dimensionless momentum of the moving trap center as the constant $P_c \equiv V/\omega_0$. We then shift to the comoving frame of the moving potential, defined by the relative coordinates $X_{\rm com}(t) \equiv X(t) - X_c(t)$ and $P_{\rm com}(t) \equiv P(t) - P_c$. Substituting our solutions into these definitions gives:
\bea
X_{\rm com}(t) &=& A \cos(\omega_0 t) + B \sin(\omega_0 t), \\
P_{\rm com}(t) &=& -A \sin(\omega_0 t) + B \cos(\omega_0 t).
\eea

By evaluating these expressions at $t=0$, we identify the constants as the initial comoving coordinates: $A = X_{\rm com}(0)$ and $B = P_{\rm com}(0)$. Thus, the comoving evolution can be reorganized into a standard rotation matrix form:
\bea
\begin{pmatrix} X_{\rm com}(t) \\ P_{\rm com}(t) \end{pmatrix}
=
\begin{pmatrix} \cos(\omega_0 t) & \sin(\omega_0 t) \\ -\sin(\omega_0 t) & \cos(\omega_0 t) \end{pmatrix}
\begin{pmatrix} X_{\rm com}(0) \\ P_{\rm com}(0) \end{pmatrix}.
\eea
This explicitly proves that, under a linear ramp potential, the state's center undergoes a perfect clockwise circular rotation at frequency $\omega_0$ around the moving center $\big(X_c(t), P_c\big)$.\\

\noindent{\bf Steps 1 \& 2: Evolution up to the second ramp ($t_2$)}

During the first ramp ($0\le t<t_{1}$), the potential moves to $R$ with dimensionless momentum $P_{c1} = R/\theta_{\rm sw}$. Starting from the origin $(0,0)$, the state rotates circularly around this moving center. Applying the rotation matrix over the angle $\theta_{\rm sw}$, the state at time $t_1$ is:
\bea
X(t_1) - R = - \sin \theta_{\rm sw} \cdot P_{c1}, \quad P(t_1) = (1 - \cos \theta_{\rm sw}) \cdot P_{c1}.
\eea

During the first wait period ($t_1\le t<t_{2}$), the potential is stationary at $R$ ($P_c = 0$). The state rotates by an angle $\theta_{\rm wait}$ around $(R, 0)$, yielding the state at time $t_2$:
\bea
X(t_2) - R &=& \cos \theta_{\rm wait} \cdot (X(t_1) - R) + \sin \theta_{\rm wait} \cdot P(t_1), \\
P(t_2) &=& - \sin \theta_{\rm wait} \cdot (X(t_1) - R) + \cos \theta_{\rm wait} \cdot P(t_1).
\eea\\

\noindent{\bf Step 3: The Symmetry Condition}

At $t_2$, the potential enters its second ramp ($t_2\le t<t_{3}$), moving backward toward $D-R$ with momentum $P_{c2} = (D-2R)/\theta_{\rm sw}$. For the protocol to successfully catch the particle in the final ground state, the trajectory must satisfy a symmetry condition. Geometrically, this means that, in the comoving frame of the second ramp, the state at $t_2$ must form an isosceles triangle with the vertical $\hat{P}$-axis. Equivalently, the angle of the relative state vector must equal half of the total rotation angle during the upcoming ramp, namely $\theta_{\rm sw}/2$. Using the relative coordinates $X(t_2) - R$ and $P(t_2) - P_{c2}$, this symmetry condition can be written as:
\bea
\frac{R - X(t_2)}{P(t_2) - P_{c2}}=\tan\left(\frac{\theta_{\rm sw}}{2}\right).
\eea\\

\noindent{\bf Step 4: Solving for the Valid Protocol}

Substituting the expressions for $X(t_2)$, $P(t_2)$, $X(t_1)$, and $P(t_1)$ into the symmetry constraint and applying trigonometric addition and half-angle identities simplifies the equation remarkably to:
\bea
- \cos(\theta_{\rm sw} + \theta_{\rm wait}) \cdot 2 P_{c1} = P_{c2}.
\eea

Substituting the momenta $P_{c1} = R/\theta_{\rm sw}$ and $P_{c2} = (D-2R)/\theta_{\rm sw}$ into this equation gives the fundamental phase relation:
\bea
\cos[\omega_0(\tau_{\rm sw} + \tau_{\rm wait})] = \frac{2R-D}{2R}.
\eea

By inverting this relation, we find the exact required holding duration:
\bea
\tau_{\rm wait} = \frac{1}{\omega_0}\cos^{-1}\left(\frac{2R-D}{2R}\right) - \tau_{\rm sw}.
\eea
Therefore, the total transport time for the BBB protocol considering finite-switching time is:
\bea
T = 3\tau_{\rm sw} + 2\tau_{\rm wait} = 2(\tau_{\rm sw} + \tau_{\rm wait}) + \tau_{\rm sw} = \tau_{\rm BBB}(\omega_0, R) + \tau_{\rm sw}.
\eea
Remarkably, replacing instantaneous shifts with continuous linear ramps simply adds a single switching duration $\tau_{\rm sw}$ to the total transport time. The fundamental time advantage of incorporating a backward-moving potential is thus highly robust against realistic bandwidth limitations.


\end{widetext}

\begin{thebibliography}{99}
\bibitem{Nielsen2000} M. A. Nielsen and I. L. Chuang. {\it Quantum Computation and Quantum Information}, Cambridge University Press (2000) .

\bibitem{Cirac1995} J. I. Cirac and P. Zoller, Quantum Computations with Cold Trapped Ions, Phys. Rev. Lett. {\bf 74}, 4091 (1995).
\bibitem{Kielpinski2002} D. Kielpinski, C. Monroe and D. Wineland, Architecture for a large-scale ion-trap quantum computer, Nature {\bf 417}, 709–711 (2002).
\bibitem{Leibfried2003} D. Leibfried, R. Blatt, C. Monroe, and D. Wineland, Quantum dynamics of single trapped ions, Rev. Mod. Phys. {\bf 75}, 281 (2003). 
\bibitem{Haffner2008} H. Häffner, C. Roos, and R. Blatt, Quantum computing with trapped ions, Phys. Rep. {\bf 469}, 155 (2008).
\bibitem{Home2009} J. P. Home, D. Hanneke, J. D. Jost, J. M. Amini, D. Leibfried, and D. J. Wineland, Complete Methods Set for Scalable Ion Trap Quantum Information Processing, Science {\bf 325}, 1227 (2009).
\bibitem{Lo2015} H.-Y. Lo, D. Kienzler, L. de Clercq, M. Marinelli, V. Negnevitsky, B. C. Keitch, and J. P. Home, Spin–motion entanglement and state diagnosis with squeezed oscillator wavepackets, Nature {\bf 521}, 336 (2015). 
\bibitem{Bluvstein2022} D. Bluvstein, H. Levine, G. Semeghini, T. T. Wang, S. Ebadi, M. Kalinowski, A. Keesling, N. Maskara, H. Pichler {\it et al.}, A quantum processor based on coherent transport of entangled atom arrays, Nature {\bf 604}, 451 (2022).
\bibitem{Graham2022} T. M. Graham, Y. Song, J. Scott, C. Poole, L. Phuttitarn, K. Jooya, P. Eichler, X. Jiang, A. Marra {\it et al.}, Multi-qubit entanglement and algorithms on a neutral-atom quantum computer, Nature {\bf 604}, 457 (2022). 
\bibitem{Jandura2022} S. Jandura and G. Pupillo, Time-Optimal Two- and Three-Qubit Gates for Rydberg Atoms, Quantum {\bf 6}, 712 (2022). 
\bibitem{Moses2023} S. A. Moses, C. H. Baldwin, M. S. Allman, R. Ancona, L. Ascarrunz, C. Barnes, J. Bartolotta, B. Bjork, P. Blanchard {\it et al.}, A Race-Track Trapped-Ion Quantum Processor, Phys. Rev. X {\bf 13}, 041052 (2023). 
\bibitem{Evered2023} S. J. Evered, D. Bluvstein, M. Kalinowski, S. Ebadi, T. Manovitz, H. Zhou, S. H. Li, A. A. Geim, T. T. Wang {\it et al.}, High-fidelity parallel entangling gates on a neutral-atom quantum computer, Nature {\bf 622}, 268 (2023). 
\bibitem{Chang2023} T. H. Chang, T. N. Wang, H. H. Jen, and Y.-C. Chen, High-fidelity Rydberg controlled-Z gates with optimized pulses, New J. phys. {\bf 25}, 123007 (2023). 
\bibitem{Bluvstein2024} D. Bluvstein, S. J. Evered, A. A. Geim, S. H. Li, H. Zhou, T. Manovitz, S. Ebadi, M. Cain, M. Kalinowski {\it et al.}, Logical quantum processor based on reconfigurable atom arrays, Nature {\bf 626}, 58 (2024). 
\bibitem{Manetsch2025} H. J. Manetsch, G.Nomura, E. Bataille, X. Lv, K. H. Leung, and M. Endres, A tweezer array with 6,100 highly coherent atomic qubits, Nature {\bf 647}, 60 (2025). 

\bibitem{Wineland1998} D. J. Wineland, C. Monroe, W. M. Itano, D. Leibfried, B. E. King, and D. M. Meekhof, J. Res, Experimental Issues in Coherent Quantum-State Manipulation of Trapped Atomic Ions, Natl. Inst. Stand. Technol. {\bf 103}, 259 (1998).
\bibitem{Parkins1999} A. S. Parkins and H. J. Kimble, Quantum state transfer between motion and light, J. Opt. B: Quantum Semiclassical Opt. {\bf 1}, 496 (1999).
\bibitem{Rowe2002} M. A. Rowe et al, Transport of quantum states and separation of ions in a dual RF ion trap. Quant. Inform. Comput {\bf 2}, 257–271 (2002).
\bibitem{Reichle2006} R. Reichle, D. Leibfried, R.B. Blakestad, J. Britton, J.D. Jost, E. Knill, C. Langer, R. Ozeri, S. Seidelin, D.J. Wineland, Transport dynamics of single ions in segmented microstructured Paul trap arrays,  Fortschr. Phys. {\bf 54}, 666 (2006).
\bibitem{Bowler2012} R. Bowler, et al, Coherent diabatic ion transport and separation in a multizone trap array. Phys. Rev. Lett. {\bf 109}, 080502 (2012).
\bibitem{Odelin2019} D. Guéry-Odelin, A. Ruschhaupt, A. Kiely, E. Torrontegui, S. Mart\'{\i}nez-Garaot, and J. G. Muga, Shortcuts to adiabaticity: Concepts, methods, and applications, Rev. Mod. Phys. {\bf 91}, 045001 (2019).

\bibitem{Chen2010} X. Chen, A. Ruschhaupt, S. Schmidt, A. del Campo, D. Guéry-Odelin, and J. G. Muga, Fast Optimal Frictionless Atom Cooling in Harmonic Traps: Shortcut to Adiabaticity, Phys. Rev. Lett. {\bf 104}, 063002 (2010).  
\bibitem{An2016} S. An, D. Lv, A. del Campo, and K. Kim, Shortcuts to adiabaticity by counterdiabatic driving for trapped-ion displacement in phase space, Nat. Commun. {\bf 7}, 12999 (2016). 
\bibitem{Kaufmann2018} P. Kaufmann, T. F. Gloger, D. Kaufmann, M.Johanning, and C. Wunderlich, High-Fidelity Preservation of Quantum Information During Trapped-Ion Transport, Phys. Rev. Lett. {\bf 120}, 010501 (2018). 
\bibitem{Finzgar2025} J. R. Finžgar, S. Notarnicola , M. Cain , M. D. Lukin, and D. Sels, Counterdiabatic Driving with Performance Guarantees, Phys. Rev. Lett. {\bf 135}, 180602 (2025). 

\bibitem{Murphy2009} M. Murphy, L. Jiang, N. Khaneja, and T. Calarco, High-Fidelity Fast Quantum Transport with Imperfect Controls, Phys. Rev. A {\bf 79}, 020301(R) (2009).
\bibitem{Lam2021} M. R. Lam, N. Peter, T.n Groh, W.g Alt, C. Robens, D. Meschede, A. Negretti, S. Montangero, T. Calarco {\it et al.}, Demonstration of Quantum Brachistochrones between Distant States of an Atom, Phys. Rev. X {\bf 11}, 011035 (2021). 
\bibitem{Hwang2025} S. Hwang, H. Hwang, K. Kim, A. Byun, K. Kim, S. Jeong, M. P. Soegianto, and J. Ahn, Fast and reliable atom transport by optical tweezers, Optica Quantum {\bf 3}, 64 (2025).

\bibitem{Agarwal1971} G. S. Agarwal, Brownian Motion of a Quantum Oscillator, Phys. Rev. A {\bf 4}, 739 (1971).
\bibitem{Daniel1989} D. J. Daniel and G. J. Milburn, Destruction of quantum coherence in a nonlinear oscillator via attenuation and amplification, Phys. Rev. A {\bf 39}, 4628 (1989).
\bibitem{Wineland1998_1} D. J. Wineland, C. Monroe, W. M. Itano, D. Leibfried, B. E. King, D. M. Meekhof, Experimental Issues in Coherent Quantum-State Manipulation of Trapped Atomic Ions, J. Res. Natl. Inst. Stand. Technol. {\bf 103}, 259 (1998).
\bibitem{Zhang2024} Z. Zhang, M. Yuan, B. Sundar, and K. R. A. Hazzard, Motional decoherence in ultracold-Rydberg-atom quantum simulators of spin models, Phys. Rev. A {\bf 110}, 053321 (2024).

\bibitem{Wineland1998_2} D. J. Wineland, C. Monroe, D. M. Meekhof, B. E. King, D. Leibfried, W. M. Itano, J. C. Bergquist, D. Berkeland, J. J. Bollinger and J. Miller, Quantum state manipulation of trapped atomic ions, Proc. Roy. Soc. A {\bf 454}, 1969 (1998).

\bibitem{Xi2011} X. Chen, E. Torrontegui, D. Stefanatos, J.-S. Li, and J. G. Muga, Optimal trajectories for efficient atomic transport without final excitation, Phys. Rev. A {\bf 84}, 043415, (2011).
\bibitem{Torrontegui2011} E. Torrontegui, S. Ibáñez, X. Chen, A. Ruschhaupt, D. Gu\'{e}ry-Odelin, and J. G. Muga, Fast Atomic Transport without Vibrational Heating, Phys. Rev. A {\bf 83}, 013415 (2011).
\bibitem{Ding2020} Y. Ding, T.-Y. Huang, K. Paul, M. Hao, and X. Chen, Smooth bang-bang shortcuts to adiabaticity for atomic transport in a moving harmonic trap, Phys. Rev. A {\bf 101}, 063410, (2020).

\bibitem{Alonso2013} J. Alonso, F. M. Leupold, B. C. Keitch and J. P. Home, Quantum control of the motional states of trapped ions through fast switching of trapping potentials, New J. Phys. {\bf 15} 023001 (2013).
\bibitem{Viola1999} L. Viola, E. Knill and S. Lloyd, Dynamical decoupling of open quantum systems. Phys. Rev. Lett. {\bf 82}, 2417–2421 (1999).
\bibitem{Morton2006} J. J. Morton, et al, Bang-bang control of fullerene qubits using ultrafast phase gates. Nat. Phys. {\bf 2}, 40–43 (2006).
\bibitem{Alonso2016} J. Alonso, F. M. Leupold, Z. U. Solèr, M. Fadel, M. Marinelli, B. C. Keitch, V. Negnevitsky and J.P. Home, Generation of large coherent states by bang–bang control of a trapped-ion oscillator, Nat Commun {\bf 7}, 11243 (2016).
\bibitem{Damodarakurup2009} S. Damodarakurup, M. Lucamarini, G. Di Giuseppe, D. Vitali and P. Tombesi, Experimental inhibition of decoherence on flying qubits via ‘bang-bang’ control. Phys. Rev. Lett. {\bf 103}, 040502 (2009).

\bibitem{Walls1983} D. F. Walls, Squeezed states of light, Nature {\bf 306}, 141 (1983). 

\bibitem{Sudarshan1963} E. C. G. Sudarshan, Equivalence of Semiclassical and Quantum Mechanical Descriptions of Statistical Light Beams, Phys. Rev. Lett. {\bf 10}, 277 (1963). 
\bibitem{Glauber1963} R. J. Glauber, Coherent and Incoherent States of the Radiation Field, Phys. Rev. {\bf 131}, 2766 (1963). 
\bibitem{Zhang1990} W.-M. Zhang, D. H. Feng, and R. Gilmore, Coherent states: Theory and some applications, Rev. Mod. Phys. {\bf 62}, 867 (1990). 

\bibitem{Wigner1932} E. Wigner, On the Quantum Correction for Thermodynamic Equilibrium, Phys.  Rev. {\bf 40}, 749 (1932).
\bibitem{Scully1997} M. O. Scully and M. S. Zubairy. {\it Quantum Optics}, Cambridge University Press (1997).

\bibitem{SM} See supplemental material for details of the evolution of a coherent state under a series of potential shifts, analysis of asymmetric BBB protocols, anharmonic corrections, the quantum speed limit, and the finite switching time analysis for BBB protocols.

\bibitem{Mandelstam1945} L. Mandelstam and I. Tamm, The Uncertainty Relation between Energy and Time in Non-Relativistic Quantum Mechanics, J. Phys. (Moscow) {\bf 9}, 249 (1945).
\bibitem{Margolus1998} N. Margolus and L. B. Levitin, The maximum speed of dynamical evolution, Physica (Amsterdam) {\bf 120D}, 188 (1998).
\bibitem{Ness2022} G. Ness , A. Alberti and Y. Sagi, Quantum Speed Limit for States with a Bounded Energy Spectrum, Phys. Rev. Lett. {\bf 129}, 140403 (2022).

\bibitem{Beugnon2007} J. Beugnon, C. Tuchendler, H. Marion, A. Gaëtan, Y. Miroshnychenko, Y. R. P. Sortais, A. M. Lance, M. P. A. Jones, G. Messin {\it et al.}, Two-dimensional transport and transfer of a single atomic qubit in optical tweezers, Nat. Phys. {\bf 3}, 696 (2007). 

\bibitem{EM} See end matter for the calculation of the orientation angle for the squeezed coherent state and the derivation of the required potential shift for the DSBBB protocols.

\bibitem{Brownnutt2015} M. Brownnutt, M. Kumph, P. Rabl, and R. Blatt, Ion-trap measurements of electric-field noise near surfaces, Rev. Mod. Phys. {\bf 87}, 1419 (2015).

\bibitem{Kaufman2012} A. M. Kaufman, B. J. Lester, and C. A. Regal, Cooling a Single Atom in an Optical Tweezer to Its Quantum Ground State, Phys. Rev. X {\bf 2}, 041014 (2012).
\bibitem{Barredo2016} D. Barredo, S. de Léséleuc, V. Lienhard, T. Lahaye, and A. Browaeys, An atom-by-atom assembler of defect-free arbitrary two-dimensional atomic arrays, Science {\bf 354}, 1021 (2016). 
\bibitem{Barredo2018} D. Barredo, V. Lienhard, S. de Léséleuc, T. Lahaye, and A. Browaeys, Synthetic three-dimensional atomic structures assembled atom by atom, Nature {\bf 561}, 79 (2018). 
\bibitem{Brown2019} M. O. Brown, T. Thiele, C. Kiehl, T.-W. Hsu, and C. A. Regal, Gray-Molasses Optical-Tweezer Loading: Controlling Collisions for Scaling Atom-Array Assembly, Phys. Rev. X {\bf 9}, 011057 (2019).  
\bibitem{Ebadi2021} S. Ebadi, T. T. Wang, H. Levine, A. Keesling, G. Semeghini, A. Omran, D. Bluvstein, R. Samajdar {\it et al.}, Quantum phases of matter on a 256-atom programmable quantum simulator, Nature {\bf 595}, 227 (2021). 
\bibitem{Scholl2021} P. Scholl, M. Schuler, H. J. Williams, A. A. Eberharter, D. Barredo, K.-N. Schymik, V. Lienhard, L.-P. Henry, T. C. Lang {\it et al.}, Quantum simulation of 2D antiferromagnets with hundreds of Rydberg atoms, Nature {\bf 595}, 233 (2021)
\bibitem{Dordevic2021} T. Dordevi\ifmmode \acute{c}\else \'{c}\fi{}, P. Samutpraphoot, P. L. Ocola, H. Bernien, B. Grinkemeyer, I. Dimitrova, V. Vuleti\'{c}, and M. D. Lukin, Entanglement transport and a nanophotonic interface for atoms in optical tweezers, Science {\bf 373}, 1511 (2021). 

\bibitem{Caneva2009} T. Caneva, M. Murphy, T. Calarco, R. Fazio, S. Montangero, V. Giovannetti, and G. E. Santoro, Optimal control at the quantum speed limit, Phys. Rev. Lett. {\bf 103}, 240501 (2009).
\bibitem{Furst2014} H. A. Fürst, M. H. Goerz, U. G. Poschinger, M. Murphy, S. Montangero, T. Calarco, F. Schmidt-Kaler, K. Singer, and C. P. Koch, Controlling the transport of an ion: Classical and quantum mechanical solutions, New J. Phys. {\bf 16}, 075007 (2014).
\bibitem{Zhang2018} X.-M. Zhang, Z.-W. Cui, X. Wang, and M.-H. Yung, Automatic spin-chain learning to explore the quantum speed limit, Phys. Rev. A {\bf 97}, 052333 (2018).
\bibitem{Pagano2024} A. Pagano, D. Jaschke, W. Weiss, and S. Montangero, Optimal control transport of neutral atoms in optical tweezers at finite temperature, Phys. Rev. Res. {\bf 6}, 033282 (2024).
\bibitem{Morandi2025} O. Morandi, S. Nicoletti, V. Gavryusev, and L.Fallani, Optimal control in phase space applied to minimal-time transfer of thermal atoms in optical traps, Phys. Rev. A {\bf 111}, 063312 (2025).
\bibitem{Cicali2025} C. Cicali, M. Calzavara, E. Cuestas, T. Calarco, R. Zeier, and F. Motzoi, Fast neutral-atom transport and transfer between optical tweezers, Phys. Rev. Appl. {\bf 24}, 024070 (2025). 
\bibitem{Serafini2009} A. Serafini, A. Retzker and M. B. Plenio, Manipulating the quantum information of the radial modes of trapped ions: linear phononics, entanglement generation, quantum state transmission and non-locality tests. New. J. Phys. {\bf 11}, 023007 (2009). 
\end{thebibliography}

\begin{thebibliography}{99}
\bibitem{Mandelstam1945} L. Mandelstam and I. Tamm, The Uncertainty Relation between Energy and Time in Non-Relativistic Quantum Mechanics, J. Phys. (Moscow) {\bf 9}, 249 (1945).
\bibitem{Margolus1998} N. Margolus and L. B. Levitin, The maximum speed of dynamical evolution, Physica (Amsterdam) {\bf 120D}, 188 (1998).

\end{thebibliography}
\end{document}